\documentclass[%
 reprint,
 amsmath,amssymb,twocolumn,
 aps,pra,showpacs
]{revtex4}

\usepackage{cases}
\usepackage{graphicx}
\usepackage{dcolumn}
\usepackage{bm}
\usepackage{color}
\usepackage{amsfonts,latexsym}
\usepackage{enumerate}
\usepackage{amsmath}
\usepackage{amssymb}
\usepackage{rotating}

\newcommand{\dif}{\mathrm{d}}

\newcommand{\Eq}[1]{Eq.~\eqref{#1}}
\newcommand{\eq}[1]{\eqref{#1}}

\newcommand{\beq}{\begin{equation}}
\newcommand{\eeq}{\end{equation}}

\newcommand{\beqa}{\begin{eqnarray}}
\newcommand{\eeqa}{\end{eqnarray}}

\newcommand{\Beqa}{\begin{eqnarray*}}
\newcommand{\Eeqa}{\end{eqnarray*}}

\newcommand{\nn}{\nonumber}

\newcommand{\vect}[1]{\mathbf{#1}}

\begin{document}
\title{Universality in $s$-wave and higher partial wave Feshbach resonances: an illustration with
a single atom near two scattering centers}
\author{Shangguo Zhu}
   \email{shangguo.zhu@gatech.edu}
\author{Shina Tan}
    \email{shina.tan@physics.gatech.edu}
    \affiliation{School of Physics, Georgia Institute of Technology, Atlanta, Georgia 30332, USA}
\date{\today}

\begin{abstract}
It is well-known that cold atoms near $s$-wave Feshbach resonances have universal properties
that are insensitive to the short-range details of the interaction. What is less
known is that atoms near higher partial wave Feshbach resonances also have remarkable universal properties.
We illustrate this with
a single atom interacting resonantly with two fixed static centers. At a Feshbach resonance point with orbital angular momentum $L\ge1$,
we find $2L+1$ shallow bound states whose energies behave like $1/R^{2L+1}$
when the distance $R$ between the two centers is large.
We then compute corrections to the binding energies due to other parameters in the effective range expansions.
For completeness we also compute the binding energies near $s$-wave Feshbach resonances, taking into account the corrections.
Afterwards we turn to the bound states at large but finite scattering volumes.
For $p$-wave and higher partial wave resonances, we derive a simple formula for the energies in terms
of a parameter called ``proximity parameter".
These results are applicable to a free atom interacting resonantly with two atoms that are localized to two lattice sites
of an optical lattice, and to one light atom interacting with two heavy ones in free space.
Modifications of the low energy physics due to the long range Van der Waals potential are also discussed.
\end{abstract}
\pacs{67.85.-d,31.15.ac,03.65.Ge}
\maketitle

\section{\label{sec:intro}Introduction}

Cold atoms with de Broglie wave lengths that greatly exceed
the range of inter-atomic potential are known to exhibit universal properties
that are mainly determined by the $s$-wave scattering length $a_0$. 
Interactions in higher partial wave channels are usually suppressed by the centrifugal barrier,
unless the atoms are near a $p$-wave~\cite{Regal2003,Zhang2004,Ticknor2004,Schunck2005,Gunter2005,Gaebler2007,Fuchs2008,Inada2008,Li2008,Deh2008,Deb2009,Goyal2010,Yamazaki2013a}, $d$-wave~\cite{Marte2002a,Werner2005,Beaufils2009}, or higher partial wave~\cite{Chin2004a,Mark2007a,Mark2007,Knoop2008} Feshbach resonances.

The universality is particularly remarkable when the atoms are near a broad~\cite{Houbiers1998a,OHara2002} $s$-wave
Feshbach resonance, near which $a_0$ is large but the effective range $r_0$ is small~\cite{Bruun2005,Pricoupenko2006}.
For example, if $a_0$ is positive and large there is a shallow two-body bound state with energy entirely determined by $a_0$ rather than the full atomic details of the interaction~\cite{Sakurai2010,Braaten2006}:
$$E\approx-\frac{\hbar^2}{2\mu a_0^2},$$
where $\mu$ is the reduced mass, and $\hbar$ is the Planck constant over $2\pi$.
For three or more atoms with large scattering length, there is another remarkable manifestation
of universality: the Efimov effect~\cite{Efimov1970a,Efimov1970}; in particular, at $a_0=\pm\infty$, there are an infinite sequence of three-body shallow bound states whose
energies form a geometric sequence with a ratio that is determined by the quantum statistics and the mass ratios of atoms ~\cite{Amado1972,Efimov1973,Braaten2006},
but not affected by the details of short-range interaction~\cite{Efimov1970a,Efimov1970,Braaten2006}.
There have been predictions of analogous effects in two-atom systems with large $s$-wave scattering length~\cite{Tan2012}.

Unlike the $s$-wave resonance, higher partial wave Feshbach resonances have not
been extensively analyzed from the perspective of universality.
There have been efforts to look for Efimov effect near $p$-wave resonances~\cite{Macek2006,Macek2007,Braaten2012},
but Nishida pointed out that Efimov effect is not possible in such a case~\cite{Nishida2012}.

In this work, we illustrate that near higher partial wave resonances,
although there is no Efimov effect in three dimensions, the concept of universality remains powerful.
By universality we mean that there are many physical properties controlled by one or two effective parameters
for the interaction, rather than by the full atomic details. Consequently different atomic species near higher partial wave Feshbach
resonances can exhibit very similar behaviors.

We study a single free atom of mass $M$ having short-range interactions with two identical static centers.
The interaction is tuned to a resonance in the $L$th partial wave channel.
Here $L$ is the quantum number for the relative orbital angular momentum between the atom
and one center. If the $L$-wave scattering volume $a_L$ is infinity, we find that there are $2L+1$ shallow
bound states with energies entirely controlled by the $L$-wave effective range $r_L$:
\beq\label{powerlaw L>=1}
E=-\frac{\hbar^2}{2M}\frac{\chi_{Lm}^{}}{(-r_L)R^{2L+1}},~~~m=-L,-L+1,\dots,L
\eeq
if $L\ge1$. Here $R$ is the distance between the two centers,
$m$ is the projection of orbital angular momentum along the line connecting the two centers,
\beq
\chi_{Lm}^{}\equiv2(2L+1)!!(2L-1)!!\left(\begin{array}{c}2L\\L+m\end{array}\right),
\eeq
and $\left(\begin{array}{c}2L\\L+m\end{array}\right)=\frac{(2L)!}{(L+m)!(L-m)!}$ are binomial coefficients.
 $a_L$ and $r_L$ are defined as parameters in the low-energy effective range
expansion for the scattering phase shifts, \Eq{effexpa}. Note that $r_L$ is negative if $L\ge1$~\cite{Pricoupenko2006a,Pricoupenko2006,Jona-Lasinio2008,Hammer2009,Hammer2010}.
We shall also study the bound states when $a_L$ is large but finite in Secs.~\ref{sec:slightly off resonance}, \ref{sec:numerical}, and \ref{sec:simple}.

Even when $a_L=\pm\infty$, there are corrections to \Eq{powerlaw L>=1} due to a nonzero radius of the interaction.
Such corrections decay more rapidly at large $R$ than the dominant term, so that at large values of $R$ \Eq{powerlaw L>=1} becomes increasingly
more accurate. We determine these corrections in Sec.~\ref{sec:resonance}.

At the $s$-wave resonance ($a_0=\pm\infty$), the problem of a single atom interacting with two centers has been
studied in Refs.~\cite{Efimov1973,Fonseca1979,Braaten2006,Marcelis2008,Nishida2008}. It was found that the shallow binding energy
\beq
E=-\frac{\hbar^2\chi_0^2}{2MR^2}
\eeq
and is independent of the effective range $r_0$ at the leading order~\cite{Efimov1973,Fonseca1979,Braaten2006,Marcelis2008,Nishida2008}. Here $\chi_0\approx0.567143$ is the real solution to the equation $\chi_0=e^{-\chi_0}$.
This inverse square dependence of the energy on $R$ qualitatively explains the presence of Efimov effect for two heavy atoms interacting resonantly with
a light atom. Within the Born-Oppenheimer approximation
the two heavy atoms are effectively interacting via a potential equal to the binding energy for the light atom. Since the potential decays like $1/R^2$
and is strong compared to the kinetic energy of the heavy atoms,
there are a sequence of three-atom bound states with energies forming a geometric sequence~\cite{Efimov1973,Fonseca1979,Braaten2006}.

For two heavy atoms interacting with a light atom near higher partial wave Feshbach resonances,
\Eq{powerlaw L>=1} shows that Efimov effect is absent, in agreement with Ref.~\cite{Nishida2012},
since the effective potential induced by the light atom
now decays like $1/R^3$ for $p$-wave resonance, $1/R^5$ for $d$-wave resonance, etc.

Despite the absence of Efimov effect, the simple power laws for the shallow bound states in \Eq{powerlaw L>=1}
illustrates the elegance of universality in higher partial wave resonances.

In Sec.~\ref{sec:simple} we derive a more general simple approximate formula, \Eq{E simple},
when $|a_L|^{1/(2L+1)}$ and $R$ are both large and finite.

One could experimentally confirm \Eq{powerlaw L>=1} and \Eq{E simple}
using a light atom and two heavy atoms,
or using a free atom having confinement induced scattering resonance~\cite{Massignan2006,Nishida2010} with two other atoms pinned to two lattice sites of an optical lattice.

The outline of the rest of this paper is as follows.

In Sec.~\ref{sec:general} we derive the general equations for the shallow bound states of an atom near two centers.

In Sec.~\ref{sec:resonance} we derive large-$R$ expansions for the bound states at resonance ($a_L=\infty$).

In Sec.~\ref{sec:slightly off resonance} we derive the $O(1/a_L)$ corrections to the wave numbers of the bound states
slightly off resonance.

In Sec.~\ref{sec:numerical} we compute the binding energies approximately when $|a_L|^{1/(2L+1)}$ and $R$ are both large
and may be comparable.

In Sec.~\ref{sec:simple} we derive a simple formula for the binding energies near a $p$-wave or higher partial wave resonance.

In Sec.~\ref{sec:VanderWaals} we consider the effects of a long range Van der Waals interaction between
the atom and each center.

In Sec.~\ref{sec:conclude} we summarize our findings, discuss some technical limitations of this work, and argue that
despite such limitations, our results will be useful in various experimental scenarios.

\section{\label{sec:general}General properties of the bound states}
We consider an atom of mass $M$ attracted by two fixed identical scattering centers that are separated by distance $R$ and located along the $z$ axis,
with Cartesian coordinates $\vect s_1=(0,0,-R/2)$ and $\vect s_2=(0,0,+R/2)$. The potential experienced by the atom
is assumed to vanish unless the atom is very close to one center. This idealized model is a good approximation for neutral atoms,
which have a long range Van der Waals potential that decays rapidly at large distance; effects of this long range potential are discussed in Sec.~\ref{sec:VanderWaals}.

Each scattering center is characterized by phase shifts $\delta_l(k)$ where $l=0,1,2,\dots$ is the orbital angular momentum of the atom
about the center and $k\equiv i\kappa$ is the wave number. The energy of the atom is
\beq\label{E}
E=\frac{\hbar^2k^2}{2M}=-\frac{\hbar^2\kappa^2}{2M}.
\eeq
For a bound state we take $\kappa>0$.

Due to the axial symmetry of the system about the $z$ axis and the parity symmetry about the origin, we may consider bound states
with a certain orbital magnetic quantum number $m$ and a certain parity $\sigma\in\{+1,-1\}$. Outside the range of potential the wave function
of the atom is
\begin{align}\label{psi form1}
\psi_{m\sigma}(\vect r)&=\sum_{l=|m|}^\infty C_l^{m\sigma}\Big[h_l^{(1)}(i\kappa r_1)Y_{lm}(\hat{\vect r}_1)\nn\\
&\mspace{105mu}+\sigma(-1)^lh_l^{(1)}(i\kappa r_2)Y_{lm}(\hat{\vect r}_2)\Big],
\end{align}
which satisfies
\beq
\psi_{m\sigma}(-\vect r)=\sigma\psi_{m\sigma}(\vect r).
\eeq
Here $\vect r$ is the position vector of the atom,
$\vect r_\alpha\equiv\vect r-\vect s_\alpha$ is the position vector of the atom relative to a center,
$\hat{\vect r}_\alpha\equiv\vect r_\alpha/r_\alpha$, $Y_{lm}$ is the spherical harmonic,
and $h_l^{(1)}(x)$ is the spherical Hankel function of the first kind. The coefficients $C_{l}^{m\sigma}$ and the parameter $\kappa$
should be chosen such that in the vicinity of a center located at $\vect s_\alpha$ ($\alpha=1,2$)
the wave function takes the form~\cite{Sakurai2010}
\beq\label{psi form2}
\psi_{m\sigma}(\vect r)=\sum_{l=|m|}^\infty \widetilde{C}_{l\alpha}^{m\sigma}\big[j_l(k r_\alpha)\cot\delta_l(k)-n_l(k r_\alpha)\big]
Y_{lm}(\hat{\vect r}_\alpha),
\eeq
where $j_l(x)$ and $n_l(x)$ are the spherical Bessel functions of the first kind and the second kind, respectively.
Because of the well-defined parity, we need to impose condition \eq{psi form2} at one center only, such as the one located at $\vect s_1$.
At $r_1<R$ we have a useful expansion:
\beq\label{center2 on center1}
h_l^{(1)}(i\kappa r_2)Y_{lm}(\hat{\vect r}_2)\!=\!(-1)^l\!\!\sum_{l'=|m|}^\infty\!\! F_{ll'}^{|m|}(\kappa R)j_{l'}(i\kappa r_1)Y_{l'm}(\hat{\vect r}_1),
\eeq
where
\begin{widetext}
\beq
F_{ll'}^m(x)=\sum_{j=0}^{\text{min}(l,l')-m}\frac{j!(2j+2m-1)!!}{(j+2m)!}\left(\begin{array}{c}l-m\\j\end{array}\right)
\left(\begin{array}{c}l'-m\\j\end{array}\right)\sqrt\frac{(2l+1)(2l'+1)(l+m)!(l'+m)!}{(l-m)!(l'-m)!}\,
\frac{h_{l+l'-m-j}^{(1)}(ix)}{(-ix)^{m+j}},
\eeq
\end{widetext}
and $\left(\begin{array}{c}n\\j\end{array}\right)=\frac{n!}{j!(n-j)!}$ is the binomial coefficient.
Substituting the identities $h_l^{(1)}(x)=j_l(x)+in_l(x)$ and \eq{center2 on center1}
into \Eq{psi form1}, and noting that $F_{ll'}^m(x)=F_{l'l}^m(x)$, we find
\begin{align}
\psi_{m\sigma}(\vect r)&=\sum_{l=|m|}^\infty\Big\{\Big[C_l^{m\sigma}+\sum_{l'=|m|}^\infty\sigma C_{l'}^{m\sigma}F_{ll'}^{|m|}(\kappa R)\Big]j_l(i\kappa r_1)\nn\\
&\mspace{70mu}+iC_l^{m\sigma}n_l(i\kappa r_1)\Big\}Y_{lm}(\hat{\vect r}_1).
\end{align}
Comparing the above formula with the required form of the expansion of the wave function in \Eq{psi form2}, we find a set of linear equations
for the unknown coefficients $C_{l}^{m\sigma}$:
\beq\label{C}
\sum_{l'=|m|}^\infty K_{ll'}^{m\sigma}(\kappa,R)C_{l'}^{m\sigma}=0,~~~~l\ge |m|,
\eeq
where
\beq\label{K}
K_{ll'}^{m\sigma}(\kappa,R)\equiv\big[1+i\cot\delta _l(i\kappa)\big]\delta_{ll'}+\sigma F_{ll'}^{|m|}(\kappa R).
\eeq
For each pair of values $(m,\sigma)$, we must adjust $\kappa$ in order to get a
nonvanishing set of coefficients $C_l^{m\sigma}$.
Each positive value of $\kappa$ corresponds to a bound state,
whose energy is given by \Eq{E}. For shallow bound states with $k=i\kappa$ close to zero,
we shall use the effective range expansion for the phase shifts $\delta_l$~\cite{Bethe1949,Madsen2002}:
\beq\label{effexpa}
k^{2l+1}\cot\delta_l(k)=-\frac{1}{a_l}+r_l\frac{k^2}{2!}+r_l'\frac{k^4}{4!}+r_l''\frac{k^6}{6!}+O(k^8).
\eeq
The coefficients $r_l'$, $r_l''$ etc are known as shape parameters.
We will use Eqs.~\eq{C} \eq{K} \eq{effexpa} to derive well-controlled expansions for
many shallow bound states in Sec.~\ref{sec:resonance}.

At a given $m$, the two parities $\sigma=\pm1$ give rise to two different states.
We can tell which one has a lower energy by a simple symmetry analysis.
It is easy to see that
\beq
\psi_{m\sigma}(x,y,-z)=\sigma(-1)^m\psi_{m\sigma}(x,y,z).
\eeq
If $\sigma(-1)^m=-1$, the wave function vanishes on the $xy$ plane, and the energy is the same
as we would obtain by imposing a hard wall potential there; thus the closer the two centers, the closer
the hard wall is to one center, and the higher the energy, hence the energy increases as we reduce the distance between
the two centers. In the language of the chemical bond theory, the state with parity $\sigma=(-1)^{m+1}$ may be called an anti-bonding state.
We have also verified numerically that for $\sigma=(-1)^m$, for which the wave function is an even function of $z$,
the energy decreases as we reduce the distance between the two centers; see Sec.~\ref{sec:numerical}. Thus
\beq\label{sigma}
\sigma=\left\{\begin{array}{ll}(-1)^m,&\text{for bonding states},\\
(-1)^{m+1},&\text{for anti-bonding states.}\end{array}\right.
\eeq
For a particular $m$, depending on the interaction between the atom and the centers
and the distance $R$ between the two centers,
the atom may have two shallow bound states one from each parity, or
a single shallow bound state with $\sigma=(-1)^m$, or no shallow bound state.
More details will be shown in Sec.~\ref{sec:numerical}.

\section{\label{sec:resonance}Bound states at resonance}

In this section we study the shallow bound states at an $L$-wave scattering resonance.
By this resonance we mean that $a_L$ diverges but the scattering volumes $a_l$ for all other partial waves (with $l\ne L$) remain finite.
We shall call the partial wave channel with orbital angular momentum $L$ the \emph{resonant channel},
and all other channels \emph{non-resonant channels}.

At large $R$ we find $2L+1$ shallow bound states with orbital magnetic quantum numbers $m=-L,-L+1,\dots,L$ and parities $\sigma=(-1)^m$.
Using Eqs.~\eq{C} \eq{K} \eq{effexpa}, we can expand $\kappa(R)$ for these states in powers of $1/R$.
It is important to bear in mind that \Eq{C} contains the effects of \emph{all} partial wave channels, not just the resonant channel.

At an $s$-wave resonance,
\begin{widetext}
\begin{align}
\label{eq:s}
\kappa&=\frac{0.567143}{R}\Big[1+\frac{0.180948 r_0}{R}+\frac{0.0688446 r_0^2}{R^2}+\frac{-2.66638 a_1+0.0332811 r_0^3-0.00485019 r_0^\prime}{R^3}\nn\\
&\quad+\frac{-0.865928 a_1 r_0+0.0181467 r_0^4-0.00544592 r_0 r_0^\prime}{R^4}\nn\\
&\quad+\frac{(-45.6556 a_2-0.473306a_1 r_0^2+0.0106396 r_0^5-0.00483849 r_0^2 r_0^\prime+0.0000520023 r_0^{\prime\prime}+0.428823a_1^2 r_1)}{R^5}
+O\Big(\frac{1}{R^6}\Big)\Big].
\end{align}
\end{widetext}
The leading order term in the above series is consistent with previous theoretical calculations~\cite{Efimov1973,Fonseca1979,Braaten2006,Marcelis2008,Nishida2008}.
The binding energy $E$ has the $-1/R^2$ dependence, which leads to the emergence of three-body Efimov states.
Moreover, the effects of the parameters $r_0$, $r_0^\prime$, $r_0''$ etc in the effective range expansion Eq.~(\ref{effexpa}) appear only in higher order terms.
Also, the effects of other partial wave channels are all higher order corrections.
In this sense, the resonant partial wave channel is dominant, in agreement with our intuition.

The results at $p,d,f$-wave resonances are listed in the following.
\begin{widetext}
\begin{enumerate}
\item
For $p$-wave resonance:
\begin{enumerate}
\item $m=0$,
\begin{align}
\label{eq:p0}
\kappa&=\frac{\sqrt{12}}{\sqrt{-r_1}\,R^{3/2}}\Big[1-\frac{a_0}{4 R}+\frac{288-96 a_0 r_1+7 a_0^2 r_1^2}{32r_1^2R^2}
-\frac{\sqrt{3} \left(36-20 a_0 r_1+5 a_0^2 r_1^2\right)}{5 (-r_1)^{5/2}R^{5/2}}\nn\\
&\mspace{130mu}+\frac{-1536-864 a_0 r_1+288 a_0^2 r_1^2-25 a_0^3 r_1^3-64 r_1 r_1^\prime}{128 r_1^3R^3}
-\frac{\sqrt{3} \left(180+924 a_0 r_1-595 a_0^2 r_1^2+70 a_0^3 r_1^3\right)}{35 (-r_1)^{7/2}R^{7/2}}\nn\\
&\mspace{130mu}+O\Big(\frac{1}{R^4}\Big)\Big];
\end{align}
\item $m=\pm1$,
\beq
\label{eq:p1}
\kappa=\frac{\sqrt{6}}{\sqrt{-r_1}\,R^{3/2}}\Big[1+\frac{3}{2r_1 R}+\frac{2\sqrt{6}}{(-r_1)^{3/2}R^{3/2}}+\frac{9}{8 r_1^2R^2}
-\frac{57 \sqrt{6}}{5 (-r_1)^{5/2}R^{5/2}}-\frac{1083+4 r_1 r_1^\prime}{16 r_1^3R^3}
+\frac{1521\sqrt{6}}{70(-r_1)^{7/2}R^{7/2}}+O\Big(\frac{1}{R^4}\Big)\Big].
\eeq
\end{enumerate}

\item
For $d$-wave resonance:
\begin{enumerate}
\item $m=0$,
\beq
\label{eq:d0}
\kappa= \frac{\sqrt{540}}{\sqrt{-r_2}\,R^{5/2}}\Big[1-\frac{a_0}{12R}+\frac{23 a_0^2}{288R^2}+\frac{-\frac{265 a_0^3}{3456}-\frac{9 a_1}{4}+\frac{15}{r_2}}{R^3}
-\frac{\sqrt{15} a_0^2}{\sqrt{-r_2}\,R^{7/2}}+\frac{\frac{12235 a_0^4}{165888}+\frac{21 a_0 a_1}{16}-\frac{65 a_0}{4 r_2}}{R^4}+O\Big(\frac{1}{R^{9/2}}\Big)\Big];
\eeq
\item $m=\pm1$,
\begin{align}
\label{eq:d1}
\kappa&=\frac{\sqrt{360}}{\sqrt{-r_2}\,R^{5/2}}\Big[1+\frac{-\frac{9 a_1}{8}+\frac{15}{r_2}}{R^3}-\frac{15 r_2^\prime}{r_2^2R^5}
+\frac{27 \left(1600-720 a_1 r_2+13 a_1^2 r_2^2\right)}{128 r_2^2R^6}-\frac{39375 a_3}{4R^7}\nn\\
&\mspace{127mu}-\frac{45 \left(36 a_1^2 r_1 r_2^2+200 r_2^\prime-9 a_1 r_2 r_2^\prime\right)}{8 r_2^3R^8}+O\Big(\frac{1}{R^9}\Big)\Big];
\end{align}
\item $m=\pm2$,
\beq
\label{eq:d2}
\kappa=\frac{\sqrt{90}}{\sqrt{-r_2}\,R^{5/2}}\Big[1+\frac{15}{2r_2 R^3}-\frac{15 r_2^\prime}{4 r_2^2R^5}+\frac{2025}{8 r_2^2R^6}-\frac{7875 a_3}{2R^7}
-\frac{540 \sqrt{10}}{(-r_2)^{5/2}R^{15/2}}-\frac{1125r'_2}{8r_2^3R^8}+O\Big(\frac{1}{R^9}\Big)\Big].
\eeq
\end{enumerate}

\item
For $f$-wave resonance:
\begin{enumerate}
\item $m=0$,
\beq
\label{eq:f0}
\kappa=\frac{\sqrt{63000}}{\sqrt{-r_3}\,R^{7/2}}\Big[1-\frac{a_0}{40R}+\frac{79 a_0^2}{3200R^2}-\frac{\frac{3121 a_0^3}{128000}+\frac{6 a_1}{5}}{R^3}
+\frac{\frac{19731 a_0^4}{819200}+\frac{57 a_0 a_1}{100}}{R^4}-\frac{3 \sqrt{70}\, a_0^2}{2\sqrt{-r_3}\,R^{9/2}}+O\Big(\frac{1}{R^5}\Big)\Big];
\eeq
\item $m=\pm1$,
\beq
\label{eq:f1}
\kappa=\frac{\sqrt{47250}}{\sqrt{-r_3}\,R^{7/2}}\Big[1-\frac{3a_1}{5R^3}+\frac{15(147-10 a_2r_3)}{2 r_3R^5}+\frac{81 a_1^2}{50R^6}-\frac{7875 r'_3}{4r_3^2R^7}
+\frac{9 a_1 (-1407+50 a_2 r_3)}{2 r_3 R^8}+O\Big(\frac{1}{R^9}\Big)\Big];
\eeq
\item $m=\pm2$,
\begin{align}
\label{eq:f2}
\kappa&=\frac{\sqrt{18900}}{\sqrt{-r_3}\,R^{7/2}}\Big[1+\frac{15(168-5a_2r_3)}{4r_3R^5}-\frac{1575 r_3^\prime}{2 r_3^2R^7}-\frac{4862025 a_4}{4R^9}
+\frac{225 \left(155232-11760 a_2 r_3+95 a_2^2 r_3^2\right)}{32 r_3^2R^{10}}\nn\\
&\mspace{130mu}-\frac{366735600 a_5}{R^{11}}+O\Big(\frac{1}{R^{12}}\Big)\Big];
\end{align}
\item $m=\pm3$,
\beq
\label{eq:f3}
\kappa=\frac{\sqrt{3150}}{\sqrt{-r_3}\,R^{7/2}}\Big[1+\frac{315}{2 r_3 R^5}-\frac{525 r_3^\prime}{4 r_3^2R^7}-\frac{694575 a_4}{2R^9}+\frac{628425}{8 r_3^2R^{10}}
-\frac{137525850 a_5}{R^{11}}+O\Big(\frac{1}{R^{12}}\Big)\Big].
\eeq
\end{enumerate}

\end{enumerate}
\end{widetext}

From Eq.~(\ref{eq:p0})-(\ref{eq:f3}), we see that at the $L$-wave resonance, the leading order term in the expansion of $\kappa$
is proportional to $\frac{1}{\sqrt{-r_L}R^{(2L+1)/2}}$. The first order correction is due to the partial wave channels
with orbital angular momentum quantum numbers equal to $L$ or $|m|$, while the higher order corrections are also influenced by other partial wave channels.
By expanding $\kappa$ to high powers of $1/R$, one will see the effects of all the effective range expansion parameters in all the channels
that satisfy $l\ge|m|$.

In general, the leading order terms for the binding energies at the $L$-wave resonance are
\begin{equation}
\label{eq:powerlaw}
E=\begin{cases}
-\frac{\hbar^2}{2M}\frac{\chi_0^2}{R^2}, &L= 0,\\
-\frac{\hbar^2}{2M}\frac{\chi_{Lm}}{(-r_L)R^{2L+1}}, &L\geq 1,
\end{cases}
\end{equation}
where $\chi_0\approx0.567143$ is the real solution to the equation $\chi_0=e^{-\chi_0}$, $-L\le m\le L$, and
\beq
\chi_{Lm}^{}=\frac{2\,(2L+1)!!(2L-1)!!(2L)!}{(L-m)!(L+m)!}.
\eeq
To obtain \Eq{eq:powerlaw} at $L\ge1$, we used the following expansion at small $x$:
\beq\label{F small x}
F_{LL}^m(x)\!=\!(-1)^{L-m+1}\frac{\chi_{Lm}^{}}{2\,x^{2L+1}}+O(x^{-2L}).
\eeq

When two heavy particles interact resonantly with a light particle by short-range potential,
within the Born-Oppenheimer approximation they experience an effective potential equal to the binding energy $E$
for the light particle.
At an $s$-wave resonance, the $-1/R^2$ dependence of the effective potential results in the appearance of three-body Efimov states~\cite{Fonseca1979}.
At higher partial wave resonances ($L\ge1$), however, the three-body Efimov effect seems unlikely,
because the Schr\"{o}dinger equation for two heavy particles with effective potential $\propto-1/R^{2L+1}$ has no scaling symmetry.

\section{\label{sec:slightly off resonance}Corrections to $\kappa$ slightly off resonance}
Since experimentally the scattering volume $a_L$ can not be tuned exactly to infinity,
we would like to know the corrections to the expansions in Eqs.~(\ref{eq:s})-(\ref{eq:f3})
when $a_L$ is large but finite. We use Eqs.~\eq{C} \eq{K}  \eq{effexpa} to calculate these corrections
to the leading order in $1/a_L$. Our results are listed in Table~\ref{tab:corr}. In general, in the very vicinity of a resonance
\beq
\kappa\approx
\begin{cases}
\kappa\big|_\text{resonance}+\frac{1}{(1+\chi_0)a_0},&L=0,\\
\kappa\big|_\text{resonance}+\frac{R^{L+1/2}}{\sqrt{-\chi_{Lm}^{}r_L}\,a_L},&L\ge1.
\end{cases}
\eeq
\begin{table}
\begin{ruledtabular}
\renewcommand{\arraystretch}{2}
\begin{tabular}{ccccc}
        &$s$  &$p$  &$d$  &$f$  \\
\hline
  $m=0$ &$\frac{0.638104}{a_0}$   &$\frac{R^{3/2}}{2 \sqrt{3} \sqrt{-r_1} a_1}$   &$\frac{R^{5/2}}{6 \sqrt{15} \sqrt{-r_2}a_2} $   &$\frac{ R^{7/2}}{ 30 \sqrt{70}\sqrt{-r_3}a_3}$   \\
  $m=\pm1$ &   &$\frac{R^{3/2}}{ \sqrt{6}\sqrt{-r_1}a_1} $   &$\frac{R^{5/2}}{6 \sqrt{10}\sqrt{-r_2}a_2}$   &$\frac{R^{7/2}}{15 \sqrt{210}\sqrt{-r_3}a_3}$   \\
  $m=\pm2$ &   &   &$\frac{R^{5/2}}{3 \sqrt{10}\sqrt{-r_2}a_2}$   &$\frac{R^{7/2}}{30 \sqrt{21}\sqrt{-r_3}a_3}$   \\
  $m=\pm3$ &   &   &   &$\frac{R^{7/2}}{15 \sqrt{14}\sqrt{-r_3}a_3}$   \\
\end{tabular}
\end{ruledtabular}
\caption{\label{tab:corr}The $O(1/a_L)$ corrections to $\kappa$ slightly away from resonance.}
\end{table}

To guarantee the applicability of the expansions in Eqs.~(\ref{eq:s})-(\ref{eq:f3}), the above corrections must be small compared to the leading order term for $\kappa$.
This condition is met if $|a_L|^{1/(2L+1)}\gg R$.
Moreover, the effective range expansion Eq.~(\ref{effexpa}) requires $|k r_e| \ll 1$, where $r_e$ is the radius of the short-range potential
between the atom and a scattering center.
In many systems $r_e$ can be characterized by the effective range $r_L$, namely $r_e\sim |r_L|^{1/(-2L+1)}$.
For such systems we obtain the domain of applicability of the expansions in Eqs.~(\ref{eq:s})-(\ref{eq:f3}):
\begin{equation}
|r_L|^{1/(-2L+1)}\ll R\ll |a_L|^{1/(2L+1)}.
\end{equation}

\section{\label{sec:numerical}Numerical results for the binding energies near a resonance}
\begin{figure*}
\includegraphics[scale=0.5]{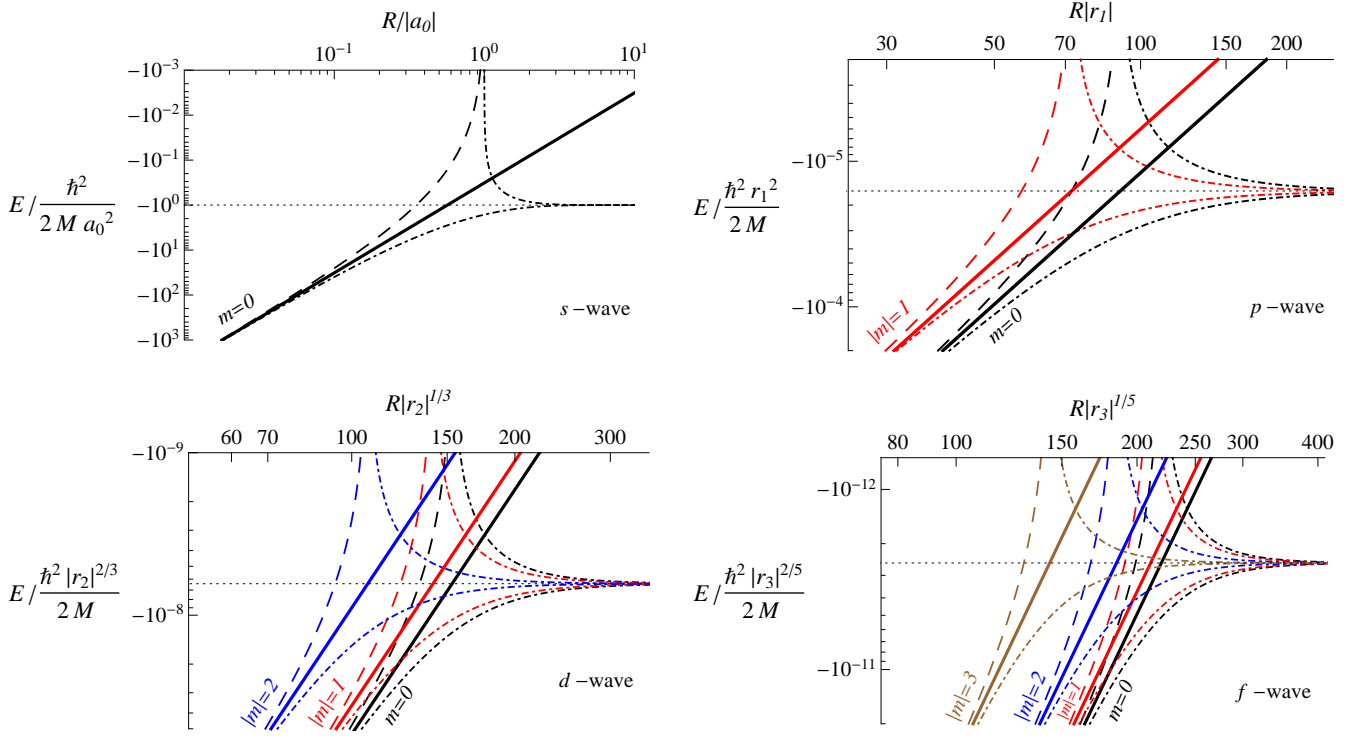}
\caption{\label{fig:eb} (color online) The binding energy $E$ of a single atom interacting resonantly with two scattering centers separated by a large distance $R$,
computed using \Eq{E} and \Eq{kappa 1 channel}.
Solid lines are for scattering volumes $a_L=\pm\infty$. Dashed lines: $a_L$ is large and negative. Dot-dashed lines: $a_L$ is large and positive.
The black, red, blue, and brown curves represent orbital magnetic quantum numbers $m=0$, $m=\pm1$, $m=\pm2$, and $m=\pm3$, respectively.
The horizontal dotted line indicates the binding energy $E_1$ due to a single scattering center [see \Eq{E1}].
We have chosen $|a_L|^{1/(2L+1)}/|r_L|^{1/(-2L+1)}=50$ or $\infty$ in the plots for $L\ge1$.}
\end{figure*}

When $R$ and $|a_L|^{1/(2L+1)}$ are both much larger than the radius $r_e$ of the potential, but $|a_L|^{1/(2L+1)}$ is comparable to $R$,
we can no longer use series expansions to compute the binding energies. Nevertheless, we can
ignore the effects of the non-resonant channels (with $l\ne L$) in \Eq{C} to a good approximation. Keeping the first two terms in the effective range
expansion for $k^{2L+1}\cot\delta_L(k)$, and noting that $k=i\kappa$, we get
\beq\label{kappa 1 channel}
\frac{1}{a_L}+\frac{r_L}{2}\kappa^{2}+(-1)^{L+1}\kappa^{2L+1}\big[1+\sigma F_{LL}^m(\kappa R)\big]=0.
\eeq
We use the above equation together with \Eq{E} to calculate the binding energies
versus $R$ for three fixed values of $a_L$
(large and positive, infinity, or large and negative) for each value of $L$.
Our results are plotted in Fig.~\ref{fig:eb}.
For the $s$-wave resonance ($L=0$), we have neglected the effective range $r_0$;
this is appropriate for broad $s$-wave Feshbach resonances~\cite{Bruun2005,Pricoupenko2006}.
For $p$-wave and higher partial wave resonances ($L\ge1$), however, the effective range $r_L$ is important~\cite{Pricoupenko2006,Macek2007,Jona-Lasinio2008,Braaten2012}
and we have kept it in our calculations.

In Fig.~\ref{fig:eb} the power laws of \Eq{eq:powerlaw} serve as asymptotes of the binding energies in the region $R\ll |a_L|^{1/(2L+1)}$.

\begin{figure*}
\includegraphics[scale=0.5]{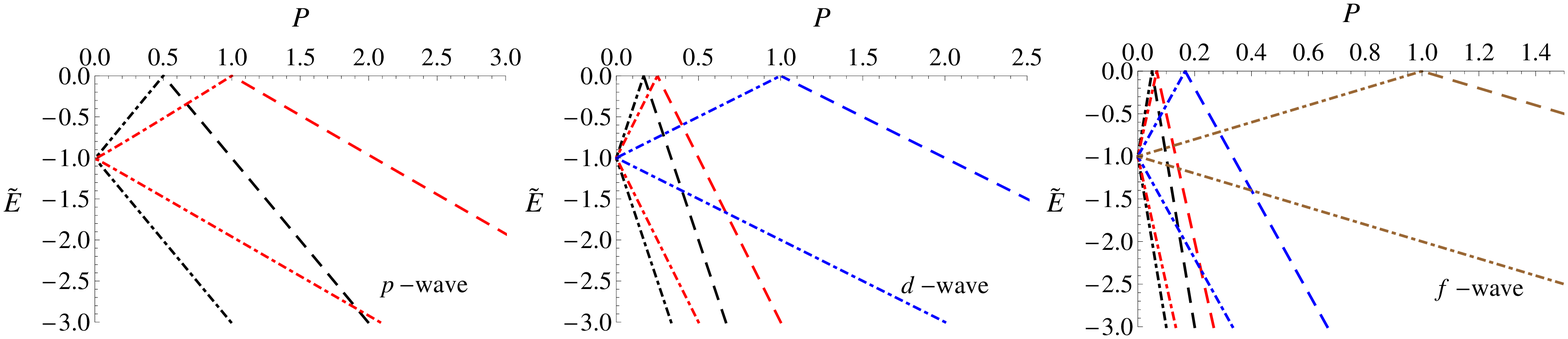}
\caption{\label{fig:EtildeP} (color online)
The dimensionless energy $\widetilde{E}=E/|E_1|$, where $E_1$ is defined in \Eq{E1}, versus the parameter $P=(2L+1)!!(2L-1)!!|a_L|/R^{2L+1}$
near $p$, $d$, or $f$-wave resonance. Curves were computed by solving \Eq{kappa 1 channel} numerically, but they have no visible differences from
the predictions of the simple formula in \Eq{E simple}. Curves with vertical intercept $-1$ are for $a_L>0$ and those with (extrapolated) vertical intercept $+1$ are for $a_L<0$.
Curves with negative slopes are for bonding states and those with positive slopes are for anti-bonding states.
Different colors indicate different values of $m$ as in Fig.~\ref{fig:eb}; in each graph, the curves with steeper slopes have smaller values of $|m|$.
}
\end{figure*}

If $a_L$ is large and negative or equal to infinity, we have a single shallow bound state for each orbital magnetic quantum number $m$
ranging from $-L$ to $L$. It is the ``bonding state" discussed in Sec.~\ref{sec:general}, and its parity is $\sigma=(-1)^m$.
If $a_L$ is large and negative, this bound state disappears when the distance between the two scattering centers exceeds a critical value,
\beq\label{Rc}
R_c=\Big(\frac{\chi_{Lm}}{2}|a_L|\Big)^{1/(2L+1)}.
\eeq
For the $s$-wave resonance, we recover the known result $R_c=|a_0|$~\cite{Yin2011}.
For the $p$-wave or any higher partial wave resonance, \Eq{Rc}
shows that the bound states with different values of $|m|$ disappear at different critical distances.
The binding energy vanishes either quadratically or linearly when $R$ approaches $R_c$ from below, depending
on $L$:
\begin{equation}
\label{eq:vanishing}
E\approx
\begin{cases}
-\frac{\hbar^2}{8Ma_0^4}(|a_0|-R)^2,&\text{~} L= 0,\\
-\frac{(2L+1)\hbar^2}{MR_c|a_Lr_L|}(R_c-R),&\text{~} L\geq 1.
\end{cases}
\end{equation}

If $a_L$ is large and positive, we have either one or two shallow bound states for each orbital magnetic quantum number $m$
ranging from $-L$ to $L$, depending on the distance $R$ between the two centers.
For all $R$ there is a ``bonding state" with parity $\sigma=(-1)^m$.
If $R>R_c$ where $R_c$ is given by \Eq{Rc} there is an extra shallow state with parity $\sigma=(-1)^{m+1}$, namely the ``anti-bonding" state
discussed in Sec.~\ref{sec:general}.
When the two scattering centers are far apart ($R\to\infty$),
the wave functions of both shallow states are localized near two separate centers,
and their energies both exponentially approach the binding energy due to a single center:
\beq\label{E1}
\renewcommand{\arraystretch}{1.6}
E_1=\left\{\begin{array}{ll}
-\frac{\hbar^2}{2Ma_0^2},&L=0,\\
-\frac{\hbar^2}{M|a_Lr_L|},&L\ge1.\end{array}\right.
\eeq
When $R$ approaches  $R_c$ from above, the energy of the anti-bonding state
vanishes linearly:
\beq
E\approx\left\{\begin{array}{ll}
-\frac{\hbar^2}{Ma_0^3}(R-a_0),&L=0,\\
-\frac{(2L+1)\hbar^2}{MR_c|a_Lr_L|}(R-R_c),&L\ge1.\end{array}\right.
\eeq

\section{\label{sec:simple}A simple formula for the binding energies near $p$-wave or higher partial wave resonances}

Near a $p$-wave or higher partial wave resonance, if $R,|a_L|^{1/(2L+1)}\gg |r_L|^{1/(1-2L)}$,
the shallow bound states satisfy $\kappa R\ll1$. Thus
\begin{align}
&(-1)^{L+1}\kappa^{2L+1}\big[1+\sigma F_{LL}^m(\kappa R)\big]\nn\\
&\quad\quad=\frac{\chi_{Lm}}{2}\sigma_zR^{-2L-1}+O(R^{-2L+1}\kappa^2),~~~L\ge1,
\end{align}
where $\sigma_z\equiv\sigma/(-1)^m$ is the $z$-parity of the wave function; $\sigma_z$ equals $1$ for bonding states and $-1$ for anti-bonding states
[see \Eq{sigma}].
Further using the inequality $R^{-2L+1}\kappa^2\ll|r_L|\kappa^2$ and \Eq{E},
we can approximate \Eq{kappa 1 channel} as
\beq
\frac{1}{a_L}-\frac{r_LME}{\hbar^2}+\frac{\chi_{Lm}}{2}\sigma_zR^{-2L-1}=0~~~\text{if $L\ge1$}.
\eeq
From the above equation we obtain a simple formula for the binding energy at $L\ge1$:
\beq\label{E simple}
\widetilde{E}=-\mathrm{sign}(a_L)-\sigma_z\left(\begin{array}{c}2L\\L-m\end{array}\right)P,
\eeq
where
\beq\label{Etilde}
\widetilde{E}\equiv\frac{E}{|E_1|},~~~P\equiv\frac{(2L+1)!!(2L-1)!!|a_{L}|}{R^{2L+1}}.
\eeq
$E_1$ is the binding energy due to a single scattering center with scattering volume $|a_L|$, and is given in \Eq{E1}.
We will refer to the parameter $P$ as the ``proximity parameter". For a given separation between the two scattering centers, if we tune the interaction
closer to resonance, the proximity parameter becomes larger.
For a given scattering volume, if we place the two scattering centers closer, the proximity parameter gets larger.

Equation \eq{E simple} illustrates the concept of universality in $p$-wave and higher partial wave Feshbach resonances.
By normalizing the binding energy $E$ by the energy due to a single scattering center with a positive scattering volume,
and normalizing the distance by the scattering volume, we get the \emph{same} set of universal relations between the dimensionless parameters
$\widetilde E$ and $P$ regardless of the atomic species or the Feshbach resonance concerned.
When we plot $\widetilde E$ against $P$, we shall find a set of \emph{straight lines}
with intercept $-1$ for $a_L<0$ or $+1$ for $a_L>0$; the absolute values of their slopes are given by the binomial coefficients
$\left(\begin{array}{c}2L\\L-m\end{array}\right)=\left(\begin{array}{c}2L\\L+m\end{array}\right)$, with the lines for $m=0$
having the steepest slopes for a given $L$.
For example, for the $d$-wave resonance, the absolute values of these slopes have ratio $1:4:6:4:1$.

In Fig.~\ref{fig:EtildeP} we re-plot the energies obtained numerically from \Eq{kappa 1 channel}.
For the parameters we used [see the caption of Fig.~\ref{fig:eb}],
we find that the resultant curves have no visible differences from the predictions of \Eq{E simple}.

\section{\label{sec:VanderWaals}Effects of the Van der Waals potential}
In the previous sections we assumed that the interaction potential between the atom and each scattering center
strictly vanishes beyond a certain radius $r_e$. If each scattering center is a pinned neutral atom and interacts with the free atom
by a long range Van der Waals potential
\beq\label{VanderWaals}
V(r)=-\frac{\hbar^2\beta_6^4}{2Mr^6},~~~r\gg r_e,
\eeq
Eq.~\eq{powerlaw L>=1} as well as the more general universal formula \Eq{E simple}
\emph{remain valid} near $p$-wave and $d$-wave resonances ($L=1,2$), but break down for $L\ge3$.
The validity of Eqs.~\eq{powerlaw L>=1} \eq{E simple} depends on the validity
of the effective range expansion, Eq.~\eqref{effexpa}.
For neutral atoms with a long range Van der Waals potential, the effective range expansion
remains valid up to the first two terms on the right hand side of Eq.~\eqref{effexpa} in the $s$-wave channel~\cite{Levy1963,Gao1998a}.
Near or at a $p$-wave resonance, Gao's work \cite{Gao1998a,Gao2009} implies that
the $p$-wave scattering length and effective range can also be defined~\cite{Gao1998a, Gao2009, Hammer2010}.
Ref. \cite{Zhang2010a} discussed the conditions under which the effective range expansion remains valid.

Near a $d$-wave resonance, Gao's work \cite{Gao2009,Gao2011} also implies that 
the first two terms of the effective range expansion for the $d$-wave phase shift are applicable within a certain window of collision energies.
We have confirmed this by doing some simple perturbative expansions of the wave function
(see Appendix~\ref{apdx:dwavephaseshift} for details). In particular, we got
\begin{equation}\label{phaseshift2}
\tan\delta_2(k)=\frac{\pi}{315}\tilde{k}^4-\frac{\pi}{900 \sqrt{2} \Gamma(5/4)^2}\frac{\tilde{k}^5}{[j+c \tilde{k}^2+o(\tilde{k}^2)]},
\end{equation}
where $\tilde{k}=k\beta_6$, and the parameters $j$ and $c$ are determined by the short-range physics.
Equation~\eq{phaseshift2} is consistent with Gao's previous results~\cite{Gao2009,Gao2011,connectiontoGao}.
For safety we have verified \Eq{phaseshift2} numerically by solving the Schr\"{o}dinger equation with the $-C_6/r^6$ potential
in three different cases: $j$ is zero, $j$ is small and negative, or $j$ is small and positive.
According to Eq.~\eqref{phaseshift2}, near a $d$-wave resonance (so that $j$ is small) and at $|j|\ll |\tilde{k}|\ll\frac{1}{c}$, the term $\frac{\pi}{315}\tilde{k}^4$ is negligible,
and we reproduce two terms of the usual $d$-wave effective range expansion [\Eq{effexpa}] if we set
\beq
a_2=\frac{\pi\beta_6^5}{900\sqrt{2}\Gamma(5/4)^2j},~~
r_2=-\frac{1800\sqrt{2}\Gamma(5/4)^2 c}{\pi\beta_6^3}.
\eeq



Thus, the first two terms in the effective range expansion [\Eq{effexpa}] remain applicable near $s,p,d$-wave Feshbach resonances.
Using these formulas for the phase shifts
and the approach developed in Sec.~\ref{sec:general}, we can derive \Eq{powerlaw L>=1} as well as the more general universal formula
\Eq{E simple} in the presence of the Van der Waals potential for $p$-wave or $d$-wave resonances.

\section{\label{sec:conclude}Summary and discussions}

In summary, we have investigated the shallow bound states of a single atom interacting strongly with two scattering centers separated by a large distance $R$.

At $s$, $p$, $d$, or $f$-wave resonances, we have obtained systematic large-$R$ expansions for the wave numbers of the bound states.
Effects of non-resonant partial wave channels and the shape parameters in the effective range expansion
enter as correction terms in these expansions.
For a $p$-wave or higher partial wave resonance ($L\ge1$)
the leading term for the binding energy behaves like $-1/R^{2L+1}$ which, if combined with the Born-Oppenheimer
approximation for two heavy and one light atoms, does not support three-body Efimov effect.
Refs.~\cite{Nishida2012,Braaten2012} contain more general discussions of the question of Efimov effect near higher partial wave
resonances, without using the BO approximation.

Slightly off resonance we have found the $O(1/a_L)$ corrections to the wave numbers for the bound states.

For large scattering volumes $a_L$ and large distances $R$ we have computed the energies of the shallow bound states approximately,
by ignoring the non-resonant partial wave scatterings and the shape parameters in the effective range expansion.

Finally, for $p$-wave and higher partial wave resonances,
we have found a simple formula for the binding energies, \Eq{E simple}, assuming that $R$ and $|a_L|^{1/(2L+1)}$ are both
large but may be comparable to each other.
Experimental confirmation of this formula will demonstrate the notion of universality in higher partial wave resonances.

Although we mainly considered short-range interactions between the scattering centers and the atom,
we have found that a long range Van der Waals potential will \emph{not} undermine our central results,
such as Eqs.~\eq{powerlaw L>=1} \eq{E simple}, near $p$ and $d$-wave resonances ($L=1,2$).

It is worthwhile to point out a technical limitation of our present work.
We have not taken into account the spatial anisotropy of the interaction near a $p$-wave or higher partial wave magnetically-tuned Feshbach resonance.
However, if the magnetic field is \emph{parallel} to the line connecting the two scattering centers, our results remain applicable;
for the shallow bound states with orbital magnetic quantum number $m$, one will replace $a_l$, $r_l$, etc in our formulas by $a_{lm}$ and $r_{lm}$ etc.
Moreover, if the higher partial wave resonance between the atom and each scattering center is physically a confinement-induced resonance~\cite{Massignan2006,Nishida2010}
due to an underlying $s$-wave Feshbach resonance, one can have a nearly isotropic effective interaction -- provided that a spherically symmetric trap
is used to confine each pinned atom which, in turn, serves as a scattering center.

To realize the scenario of one atom interacting with two fixed centers,
it is not necessary to use two atomic species with a large ratio of bare masses.
If two atoms of species A are pinned at two lattice sites of an optical lattice,
but an atom of species B is not attracted or repelled by the lattice, we can treat atoms A as infinitely heavy scattering centers for atom B.

\begin{acknowledgments}
We gratefully acknowledge support by the National Science Foundation under Grant No. PHY-1068511 and by the Alfred P. Sloan Foundation.
ST thanks Chris Greene and Paul Julienne for discussions.
\end{acknowledgments}

\section*{Note Added}
Near completion of this work an analogous study focused on $p$-wave resonance appeared online~\cite{Efremov2013}.

\appendix
\section{$d$-wave phase shift in the presence of a Van der Waals potential\label{apdx:dwavephaseshift}}
In this appendix we derive a formula for the $d$-wave scattering phase shift $\delta_2$ for a potential that behaves as \Eq{VanderWaals}.

At $r\gg r_e$ the $d$-wave radial Sch\"{o}dinger equation is
\begin{equation}\label{dwaveradial}
-u''(r)+\Big(\frac{6}{r^2}-\frac{\beta_6^4}{r^6}\Big)u(r)=k^2u(r),
\end{equation}
where $u(r)\equiv r\psi(r)$. In the following we set the Van der Waals length $\beta_6=1$ for convenience.

Assuming that $k\ll1$, we solve \Eq{dwaveradial} in two regions:
\begin{itemize}
  \item Region I: $r\ll 1/k$;
  \item Region II: $r\gg 1$.
\end{itemize}
Then, we require the two solutions to be the same in the intermediate region, $1\ll r\ll1/k$.
From this condition we obtain an approximate formula for the $d$-wave phase shift $\delta_2$.

In the region $r\ll 1/k$, we can treat the term $k^2u(r)$ in \Eq{dwaveradial} as a perturbation,
and expand $u(r)$ in powers of $k^2$:
\begin{equation}\label{urange1}
u(r)=f(r)+j \tilde{f}(r)+k^2[g(r)+j \tilde{g}(r)+c\tilde{f}(r)]+O(k^4)
\end{equation}
under a suitable normalization. Here
\beq
f(r)=\sqrt{r}\,J_{5/4}\Big(\frac{1}{2r^2}\Big)
\eeq
and
\beq
\tilde{f}(r)=\sqrt{r}\,J_{-5/4}\Big(\frac{1}{2r^2}\Big)
\eeq
are two independent solutions of the zero-energy radial Sch\"{o}dinger equation
\begin{align}
\hat{H}f(r)=0,
\end{align}
where $\hat{H}\equiv-\frac{\dif^2}{\dif r^2}+\frac{6}{r^2}-\frac{1}{r^6}$, and $J_n(x)$ is the Bessel function.
The functions $g(r)$, $\tilde{g}(r)$ are the solutions to the following equations
\begin{align}
\hat{H}g(r)=f(r),\\
\hat{H}\tilde{g}(r)=\tilde{f}(r),
\end{align}
subject to the conditions \Eq{gasymp} and \Eq{gtildeasymp} below.
At large $r$, the functions $f(r)$, $\tilde{f}(r)$, $g(r)$, $\tilde{g}(r)$ have the following asymptotic expansions:
\begin{align}
f(r)&=\frac{1}{5\sqrt{2}\Gamma(5/4)}\frac{1}{r^2}+O(r^{-6}),\\
\tilde{f}(r)&=\frac{4\sqrt{2}}{\Gamma (-1/4)} r^{3}+\frac{\sqrt2}{\Gamma(-1/4)r} +O(r^{-5}),\\
g(r)&=\frac{1}{30\sqrt{2} \Gamma(5/4)}+O(r^{-4}),\label{gasymp}\\
\tilde{g}(r)&=-\frac{2 \sqrt{2} }{7 \Gamma (-1/4)}r^5+\frac{5}{21\sqrt{2}\Gamma(-1/4)}r+O(r^{-3}).\label{gtildeasymp}
\end{align}
The constants $j$ and $c$ are determined by the short-range physics. At resonance, $j=0$ and the zero-energy solution for $u(r)$
decays like $r^{-2}$ at large $r$, consistent with the picture of a bound state at threshold.
Slightly away from $d$-wave resonance $j$ is close to zero and may take either sign.

In the region $r\gg1$, we can treat the term $\frac{\beta_6^4}{r^6}u(r)$ in \Eq{dwaveradial} as a perturbation,
and obtain another expansion for $u(r)$:
\begin{align}\label{urange2}
u(r)&=A(k)\bigg\{\sum_{n=0}^\infty k^{4n}V_n(kr)\nn\\
&\quad\quad\quad\quad-\big[\tan\delta_2(k)\big]\sum_{n=0}^\infty k^{4n}W_n(kr)\bigg\},
\end{align}
where $A(k)$ is a normalization factor depending on $k$ only,
\beq
V_0(\xi)=\xi j_2(\xi),~~~W_0(\xi)=\xi n_2(\xi),
\eeq
$j_2(\xi)$ and $n_2(\xi)$ are the spherical Bessel functions,
and $V_n(\xi)$ and $W_n(\xi)$ are defined by the equations
\beq
-V_n''(\xi)+\Big(\frac{6}{\xi^2}-1\Big)V_n(\xi)=\frac{1}{\xi^6}V_{n-1}(\xi),
\eeq
\beq
-W_n''(\xi)+\Big(\frac{6}{\xi^2}-1\Big)W_n(\xi)=\frac{1}{\xi^6}W_{n-1}(\xi),
\eeq
subject to the condition that $V_n(\xi)$ and $W_n(\xi)$ are oscillating \emph{decaying} functions at $\xi\to\infty$
for $n\ge1$.

When $\xi\to0^+$ we have
\begin{align}
&V_0(\xi)=\frac{\xi^3}{15}+O(\xi^5),\\
&W_0(\xi)=-\frac{3}{\xi ^2}-\frac{1}{2}-\frac{\xi ^2}{8}+O(\xi^4),\\
&V_1(\xi)=-\frac{\pi }{105 \xi^2}+O\Big(\frac{1}{\xi}\Big),\\
&W_1(\xi)=\frac{1}{12 \xi^6}+O\Big(\frac{1}{\xi^4}\Big).
\end{align}

In the intermediate region, $1\ll r\ll 1/k$, we obtain from \Eq{urange1}
\begin{align}
\label{eq:usmallk}
u(r)&\propto \big[j+c k^2+O(k^4)\big]r^{3}-\frac{\pi }{20 \sqrt{2}\Gamma (5/4)^2}\frac{1}{ r^2}\nn\\
&\quad+\text{(other powers of $r$)}
\end{align}
and obtain from \Eq{urange2}
\begin{align}
u(r)&\propto\Big[\frac{k^3}{15}+o(k^6)+O(k^7\tan\delta_2)\Big]r^3\nn\\
&\quad+\Big\{-\frac{\pi k^2}{105}+o(k^5)+\Big[\frac{3}{k^2}+o(k)\Big]\tan\delta_2\Big\}r^{-2}\nn\\
&\quad+\text{(other powers of $r$)}.
\end{align}
Note that the ``other powers of $r$" include terms of the order $r^n(\ln r)^m$ for integers $n$ and $m$,
with $n\ne-2,3$.
Comparing the above two formulas for $u(r)$ in the intermediate region, we
get the expression Eq.~\eqref{phaseshift2} for the phase shift $\delta_2$.
%

%
%


\begin{thebibliography}{54}
\expandafter\ifx\csname natexlab\endcsname\relax\def\natexlab#1{#1}\fi
\expandafter\ifx\csname bibnamefont\endcsname\relax
  \def\bibnamefont#1{#1}\fi
\expandafter\ifx\csname bibfnamefont\endcsname\relax
  \def\bibfnamefont#1{#1}\fi
\expandafter\ifx\csname citenamefont\endcsname\relax
  \def\citenamefont#1{#1}\fi
\expandafter\ifx\csname url\endcsname\relax
  \def\url#1{\texttt{#1}}\fi
\expandafter\ifx\csname urlprefix\endcsname\relax\def\urlprefix{URL }\fi
\providecommand{\bibinfo}[2]{#2}
\providecommand{\eprint}[2][]{\url{#2}}

\bibitem[{\citenamefont{Regal et~al.}(2003)\citenamefont{Regal, Ticknor, Bohn,
  and Jin}}]{Regal2003}
\bibinfo{author}{\bibfnamefont{C.~A.} \bibnamefont{Regal}},
  \bibinfo{author}{\bibfnamefont{C.}~\bibnamefont{Ticknor}},
  \bibinfo{author}{\bibfnamefont{J.~L.} \bibnamefont{Bohn}}, \bibnamefont{and}
  \bibinfo{author}{\bibfnamefont{D.~S.} \bibnamefont{Jin}},
  \bibinfo{journal}{Phys. Rev. Lett.} \textbf{\bibinfo{volume}{90}},
  \bibinfo{pages}{053201} (\bibinfo{year}{2003}).

\bibitem[{\citenamefont{Zhang et~al.}(2004)\citenamefont{Zhang, van Kempen,
  Bourdel, Khaykovich, Cubizolles, Chevy, Teichmann, Tarruell, Kokkelmans, and
  Salomon}}]{Zhang2004}
\bibinfo{author}{\bibfnamefont{J.}~\bibnamefont{Zhang}},
  \bibinfo{author}{\bibfnamefont{E.~G.~M.} \bibnamefont{van Kempen}},
  \bibinfo{author}{\bibfnamefont{T.}~\bibnamefont{Bourdel}},
  \bibinfo{author}{\bibfnamefont{L.}~\bibnamefont{Khaykovich}},
  \bibinfo{author}{\bibfnamefont{J.}~\bibnamefont{Cubizolles}},
  \bibinfo{author}{\bibfnamefont{F.}~\bibnamefont{Chevy}},
  \bibinfo{author}{\bibfnamefont{M.}~\bibnamefont{Teichmann}},
  \bibinfo{author}{\bibfnamefont{L.}~\bibnamefont{Tarruell}},
  \bibinfo{author}{\bibfnamefont{S.~J. J. M.~F.} \bibnamefont{Kokkelmans}},
  \bibnamefont{and} \bibinfo{author}{\bibfnamefont{C.}~\bibnamefont{Salomon}},
  \bibinfo{journal}{Phys. Rev. A} \textbf{\bibinfo{volume}{70}},
  \bibinfo{pages}{030702} (\bibinfo{year}{2004}).

\bibitem[{\citenamefont{Ticknor et~al.}(2004)\citenamefont{Ticknor, Regal, Jin,
  and Bohn}}]{Ticknor2004}
\bibinfo{author}{\bibfnamefont{C.}~\bibnamefont{Ticknor}},
  \bibinfo{author}{\bibfnamefont{C.~A.} \bibnamefont{Regal}},
  \bibinfo{author}{\bibfnamefont{D.~S.} \bibnamefont{Jin}}, \bibnamefont{and}
  \bibinfo{author}{\bibfnamefont{J.~L.} \bibnamefont{Bohn}},
  \bibinfo{journal}{Phys. Rev. A} \textbf{\bibinfo{volume}{69}},
  \bibinfo{pages}{042712} (\bibinfo{year}{2004}).

\bibitem[{\citenamefont{Schunck et~al.}(2005)\citenamefont{Schunck, Zwierlein,
  Stan, Raupach, Ketterle, Simoni, Tiesinga, Williams, and
  Julienne}}]{Schunck2005}
\bibinfo{author}{\bibfnamefont{C.~H.} \bibnamefont{Schunck}},
  \bibinfo{author}{\bibfnamefont{M.~W.} \bibnamefont{Zwierlein}},
  \bibinfo{author}{\bibfnamefont{C.~A.} \bibnamefont{Stan}},
  \bibinfo{author}{\bibfnamefont{S.~M.~F.} \bibnamefont{Raupach}},
  \bibinfo{author}{\bibfnamefont{W.}~\bibnamefont{Ketterle}},
  \bibinfo{author}{\bibfnamefont{A.}~\bibnamefont{Simoni}},
  \bibinfo{author}{\bibfnamefont{E.}~\bibnamefont{Tiesinga}},
  \bibinfo{author}{\bibfnamefont{C.~J.} \bibnamefont{Williams}},
  \bibnamefont{and} \bibinfo{author}{\bibfnamefont{P.~S.}
  \bibnamefont{Julienne}}, \bibinfo{journal}{Phys. Rev. A}
  \textbf{\bibinfo{volume}{71}}, \bibinfo{pages}{045601}
  (\bibinfo{year}{2005}).

\bibitem[{\citenamefont{G\"unter et~al.}(2005)\citenamefont{G\"unter,
  St\"oferle, Moritz, K\"ohl, and Esslinger}}]{Gunter2005}
\bibinfo{author}{\bibfnamefont{K.}~\bibnamefont{G\"unter}},
  \bibinfo{author}{\bibfnamefont{T.}~\bibnamefont{St\"oferle}},
  \bibinfo{author}{\bibfnamefont{H.}~\bibnamefont{Moritz}},
  \bibinfo{author}{\bibfnamefont{M.}~\bibnamefont{K\"ohl}}, \bibnamefont{and}
  \bibinfo{author}{\bibfnamefont{T.}~\bibnamefont{Esslinger}},
  \bibinfo{journal}{Phys. Rev. Lett.} \textbf{\bibinfo{volume}{95}},
  \bibinfo{pages}{230401} (\bibinfo{year}{2005}).

\bibitem[{\citenamefont{Gaebler et~al.}(2007)\citenamefont{Gaebler, Stewart,
  Bohn, and Jin}}]{Gaebler2007}
\bibinfo{author}{\bibfnamefont{J.~P.} \bibnamefont{Gaebler}},
  \bibinfo{author}{\bibfnamefont{J.~T.} \bibnamefont{Stewart}},
  \bibinfo{author}{\bibfnamefont{J.~L.} \bibnamefont{Bohn}}, \bibnamefont{and}
  \bibinfo{author}{\bibfnamefont{D.~S.} \bibnamefont{Jin}},
  \bibinfo{journal}{Phys. Rev. Lett.} \textbf{\bibinfo{volume}{98}},
  \bibinfo{pages}{200403} (\bibinfo{year}{2007}).

\bibitem[{\citenamefont{Fuchs et~al.}(2008)\citenamefont{Fuchs, Ticknor, Dyke,
  Veeravalli, Kuhnle, Rowlands, Hannaford, and Vale}}]{Fuchs2008}
\bibinfo{author}{\bibfnamefont{J.}~\bibnamefont{Fuchs}},
  \bibinfo{author}{\bibfnamefont{C.}~\bibnamefont{Ticknor}},
  \bibinfo{author}{\bibfnamefont{P.}~\bibnamefont{Dyke}},
  \bibinfo{author}{\bibfnamefont{G.}~\bibnamefont{Veeravalli}},
  \bibinfo{author}{\bibfnamefont{E.}~\bibnamefont{Kuhnle}},
  \bibinfo{author}{\bibfnamefont{W.}~\bibnamefont{Rowlands}},
  \bibinfo{author}{\bibfnamefont{P.}~\bibnamefont{Hannaford}},
  \bibnamefont{and} \bibinfo{author}{\bibfnamefont{C.~J.} \bibnamefont{Vale}},
  \bibinfo{journal}{Phys. Rev. A} \textbf{\bibinfo{volume}{77}},
  \bibinfo{pages}{053616} (\bibinfo{year}{2008}).

\bibitem[{\citenamefont{Inada et~al.}(2008)\citenamefont{Inada, Horikoshi,
  Nakajima, Kuwata-Gonokami, Ueda, and Mukaiyama}}]{Inada2008}
\bibinfo{author}{\bibfnamefont{Y.}~\bibnamefont{Inada}},
  \bibinfo{author}{\bibfnamefont{M.}~\bibnamefont{Horikoshi}},
  \bibinfo{author}{\bibfnamefont{S.}~\bibnamefont{Nakajima}},
  \bibinfo{author}{\bibfnamefont{M.}~\bibnamefont{Kuwata-Gonokami}},
  \bibinfo{author}{\bibfnamefont{M.}~\bibnamefont{Ueda}}, \bibnamefont{and}
  \bibinfo{author}{\bibfnamefont{T.}~\bibnamefont{Mukaiyama}},
  \bibinfo{journal}{Phys. Rev. Lett.} \textbf{\bibinfo{volume}{101}},
  \bibinfo{pages}{100401} (\bibinfo{year}{2008}).

\bibitem[{\citenamefont{Li et~al.}(2008)\citenamefont{Li, Singh, Tscherbul, and
  Madison}}]{Li2008}
\bibinfo{author}{\bibfnamefont{Z.}~\bibnamefont{Li}},
  \bibinfo{author}{\bibfnamefont{S.}~\bibnamefont{Singh}},
  \bibinfo{author}{\bibfnamefont{T.~V.} \bibnamefont{Tscherbul}},
  \bibnamefont{and} \bibinfo{author}{\bibfnamefont{K.~W.}
  \bibnamefont{Madison}}, \bibinfo{journal}{Phys. Rev. A}
  \textbf{\bibinfo{volume}{78}}, \bibinfo{pages}{022710}
  (\bibinfo{year}{2008}).

\bibitem[{\citenamefont{Deh et~al.}(2008)\citenamefont{Deh, Marzok, Zimmermann,
  and Courteille}}]{Deh2008}
\bibinfo{author}{\bibfnamefont{B.}~\bibnamefont{Deh}},
  \bibinfo{author}{\bibfnamefont{C.}~\bibnamefont{Marzok}},
  \bibinfo{author}{\bibfnamefont{C.}~\bibnamefont{Zimmermann}},
  \bibnamefont{and} \bibinfo{author}{\bibfnamefont{P.~W.}
  \bibnamefont{Courteille}}, \bibinfo{journal}{Phys. Rev. A}
  \textbf{\bibinfo{volume}{77}}, \bibinfo{pages}{010701}
  (\bibinfo{year}{2008}).

\bibitem[{\citenamefont{Deb and Hazra}(2009)}]{Deb2009}
\bibinfo{author}{\bibfnamefont{B.}~\bibnamefont{Deb}} \bibnamefont{and}
  \bibinfo{author}{\bibfnamefont{J.}~\bibnamefont{Hazra}},
  \bibinfo{journal}{Phys. Rev. Lett.} \textbf{\bibinfo{volume}{103}},
  \bibinfo{pages}{023201} (\bibinfo{year}{2009}).

\bibitem[{\citenamefont{Goyal et~al.}(2010)\citenamefont{Goyal, Reichenbach,
  and Deutsch}}]{Goyal2010}
\bibinfo{author}{\bibfnamefont{K.}~\bibnamefont{Goyal}},
  \bibinfo{author}{\bibfnamefont{I.}~\bibnamefont{Reichenbach}},
  \bibnamefont{and} \bibinfo{author}{\bibfnamefont{I.}~\bibnamefont{Deutsch}},
  \bibinfo{journal}{Phys. Rev. A} \textbf{\bibinfo{volume}{82}},
  \bibinfo{pages}{062704} (\bibinfo{year}{2010}).

\bibitem[{\citenamefont{Yamazaki et~al.}(2013)\citenamefont{Yamazaki, Taie,
  Sugawa, Enomoto, and Takahashi}}]{Yamazaki2013a}
\bibinfo{author}{\bibfnamefont{R.}~\bibnamefont{Yamazaki}},
  \bibinfo{author}{\bibfnamefont{S.}~\bibnamefont{Taie}},
  \bibinfo{author}{\bibfnamefont{S.}~\bibnamefont{Sugawa}},
  \bibinfo{author}{\bibfnamefont{K.}~\bibnamefont{Enomoto}}, \bibnamefont{and}
  \bibinfo{author}{\bibfnamefont{Y.}~\bibnamefont{Takahashi}},
  \bibinfo{journal}{Phys. Rev. A} \textbf{\bibinfo{volume}{87}},
  \bibinfo{pages}{010704} (\bibinfo{year}{2013}).

\bibitem[{\citenamefont{Marte et~al.}(2002)\citenamefont{Marte, Volz, Schuster,
  D\"urr, Rempe, van Kempen, and Verhaar}}]{Marte2002a}
\bibinfo{author}{\bibfnamefont{A.}~\bibnamefont{Marte}},
  \bibinfo{author}{\bibfnamefont{T.}~\bibnamefont{Volz}},
  \bibinfo{author}{\bibfnamefont{J.}~\bibnamefont{Schuster}},
  \bibinfo{author}{\bibfnamefont{S.}~\bibnamefont{D\"urr}},
  \bibinfo{author}{\bibfnamefont{G.}~\bibnamefont{Rempe}},
  \bibinfo{author}{\bibfnamefont{E.~G.~M.} \bibnamefont{van Kempen}},
  \bibnamefont{and} \bibinfo{author}{\bibfnamefont{B.~J.}
  \bibnamefont{Verhaar}}, \bibinfo{journal}{Phys. Rev. Lett.}
  \textbf{\bibinfo{volume}{89}}, \bibinfo{pages}{283202}
  (\bibinfo{year}{2002}).

\bibitem[{\citenamefont{Werner et~al.}(2005)\citenamefont{Werner, Griesmaier,
  Hensler, Stuhler, Pfau, Simoni, and Tiesinga}}]{Werner2005}
\bibinfo{author}{\bibfnamefont{J.}~\bibnamefont{Werner}},
  \bibinfo{author}{\bibfnamefont{A.}~\bibnamefont{Griesmaier}},
  \bibinfo{author}{\bibfnamefont{S.}~\bibnamefont{Hensler}},
  \bibinfo{author}{\bibfnamefont{J.}~\bibnamefont{Stuhler}},
  \bibinfo{author}{\bibfnamefont{T.}~\bibnamefont{Pfau}},
  \bibinfo{author}{\bibfnamefont{A.}~\bibnamefont{Simoni}}, \bibnamefont{and}
  \bibinfo{author}{\bibfnamefont{E.}~\bibnamefont{Tiesinga}},
  \bibinfo{journal}{Phys. Rev. Lett.} \textbf{\bibinfo{volume}{94}},
  \bibinfo{pages}{183201} (\bibinfo{year}{2005}).

\bibitem[{\citenamefont{Beaufils et~al.}(2009)\citenamefont{Beaufils,
  Crubellier, Zanon, Laburthe-Tolra, Mar\'echal, Vernac, and
  Gorceix}}]{Beaufils2009}
\bibinfo{author}{\bibfnamefont{Q.}~\bibnamefont{Beaufils}},
  \bibinfo{author}{\bibfnamefont{A.}~\bibnamefont{Crubellier}},
  \bibinfo{author}{\bibfnamefont{T.}~\bibnamefont{Zanon}},
  \bibinfo{author}{\bibfnamefont{B.}~\bibnamefont{Laburthe-Tolra}},
  \bibinfo{author}{\bibfnamefont{E.}~\bibnamefont{Mar\'echal}},
  \bibinfo{author}{\bibfnamefont{L.}~\bibnamefont{Vernac}}, \bibnamefont{and}
  \bibinfo{author}{\bibfnamefont{O.}~\bibnamefont{Gorceix}},
  \bibinfo{journal}{Phys. Rev. A} \textbf{\bibinfo{volume}{79}},
  \bibinfo{pages}{032706} (\bibinfo{year}{2009}).

\bibitem[{\citenamefont{Chin et~al.}(2004)\citenamefont{Chin,
  Vuleti\ifmmode~\acute{c}\else \'{c}\fi{}, Kerman, Chu, Tiesinga, Leo, and
  Williams}}]{Chin2004a}
\bibinfo{author}{\bibfnamefont{C.}~\bibnamefont{Chin}},
  \bibinfo{author}{\bibfnamefont{V.}~\bibnamefont{Vuleti\ifmmode~\acute{c}\else
  \'{c}\fi{}}}, \bibinfo{author}{\bibfnamefont{A.~J.} \bibnamefont{Kerman}},
  \bibinfo{author}{\bibfnamefont{S.}~\bibnamefont{Chu}},
  \bibinfo{author}{\bibfnamefont{E.}~\bibnamefont{Tiesinga}},
  \bibinfo{author}{\bibfnamefont{P.~J.} \bibnamefont{Leo}}, \bibnamefont{and}
  \bibinfo{author}{\bibfnamefont{C.~J.} \bibnamefont{Williams}},
  \bibinfo{journal}{Phys. Rev. A} \textbf{\bibinfo{volume}{70}},
  \bibinfo{pages}{032701} (\bibinfo{year}{2004}).

\bibitem[{\citenamefont{Mark et~al.}(2007{\natexlab{a}})\citenamefont{Mark,
  Ferlaino, Knoop, Danzl, Kraemer, Chin, N\"agerl, and Grimm}}]{Mark2007a}
\bibinfo{author}{\bibfnamefont{M.}~\bibnamefont{Mark}},
  \bibinfo{author}{\bibfnamefont{F.}~\bibnamefont{Ferlaino}},
  \bibinfo{author}{\bibfnamefont{S.}~\bibnamefont{Knoop}},
  \bibinfo{author}{\bibfnamefont{J.~G.} \bibnamefont{Danzl}},
  \bibinfo{author}{\bibfnamefont{T.}~\bibnamefont{Kraemer}},
  \bibinfo{author}{\bibfnamefont{C.}~\bibnamefont{Chin}},
  \bibinfo{author}{\bibfnamefont{H.-C.} \bibnamefont{N\"agerl}},
  \bibnamefont{and} \bibinfo{author}{\bibfnamefont{R.}~\bibnamefont{Grimm}},
  \bibinfo{journal}{Phys. Rev. A} \textbf{\bibinfo{volume}{76}},
  \bibinfo{pages}{042514} (\bibinfo{year}{2007}{\natexlab{a}}).

\bibitem[{\citenamefont{Mark et~al.}(2007{\natexlab{b}})\citenamefont{Mark,
  Kraemer, Waldburger, Herbig, Chin, N\"agerl, and Grimm}}]{Mark2007}
\bibinfo{author}{\bibfnamefont{M.}~\bibnamefont{Mark}},
  \bibinfo{author}{\bibfnamefont{T.}~\bibnamefont{Kraemer}},
  \bibinfo{author}{\bibfnamefont{P.}~\bibnamefont{Waldburger}},
  \bibinfo{author}{\bibfnamefont{J.}~\bibnamefont{Herbig}},
  \bibinfo{author}{\bibfnamefont{C.}~\bibnamefont{Chin}},
  \bibinfo{author}{\bibfnamefont{H.-C.} \bibnamefont{N\"agerl}},
  \bibnamefont{and} \bibinfo{author}{\bibfnamefont{R.}~\bibnamefont{Grimm}},
  \bibinfo{journal}{Phys. Rev. Lett.} \textbf{\bibinfo{volume}{99}},
  \bibinfo{pages}{113201} (\bibinfo{year}{2007}{\natexlab{b}}).

\bibitem[{\citenamefont{Knoop et~al.}(2008)\citenamefont{Knoop, Mark, Ferlaino,
  Danzl, Kraemer, N\"agerl, and Grimm}}]{Knoop2008}
\bibinfo{author}{\bibfnamefont{S.}~\bibnamefont{Knoop}},
  \bibinfo{author}{\bibfnamefont{M.}~\bibnamefont{Mark}},
  \bibinfo{author}{\bibfnamefont{F.}~\bibnamefont{Ferlaino}},
  \bibinfo{author}{\bibfnamefont{J.~G.} \bibnamefont{Danzl}},
  \bibinfo{author}{\bibfnamefont{T.}~\bibnamefont{Kraemer}},
  \bibinfo{author}{\bibfnamefont{H.-C.} \bibnamefont{N\"agerl}},
  \bibnamefont{and} \bibinfo{author}{\bibfnamefont{R.}~\bibnamefont{Grimm}},
  \bibinfo{journal}{Phys. Rev. Lett.} \textbf{\bibinfo{volume}{100}},
  \bibinfo{pages}{083002} (\bibinfo{year}{2008}).

\bibitem[{\citenamefont{Houbiers et~al.}(1998)\citenamefont{Houbiers, Stoof,
  McAlexander, and Hulet}}]{Houbiers1998a}
\bibinfo{author}{\bibfnamefont{M.}~\bibnamefont{Houbiers}},
  \bibinfo{author}{\bibfnamefont{H.~T.~C.} \bibnamefont{Stoof}},
  \bibinfo{author}{\bibfnamefont{W.~I.} \bibnamefont{McAlexander}},
  \bibnamefont{and} \bibinfo{author}{\bibfnamefont{R.~G.} \bibnamefont{Hulet}},
  \bibinfo{journal}{Phys. Rev. A} \textbf{\bibinfo{volume}{57}},
  \bibinfo{pages}{R1497} (\bibinfo{year}{1998}).

\bibitem[{\citenamefont{O'Hara et~al.}(2002)\citenamefont{O'Hara, Hemmer,
  Granade, Gehm, Thomas, Venturi, Tiesinga, and Williams}}]{OHara2002}
\bibinfo{author}{\bibfnamefont{K.~M.} \bibnamefont{O'Hara}},
  \bibinfo{author}{\bibfnamefont{S.~L.} \bibnamefont{Hemmer}},
  \bibinfo{author}{\bibfnamefont{S.~R.} \bibnamefont{Granade}},
  \bibinfo{author}{\bibfnamefont{M.~E.} \bibnamefont{Gehm}},
  \bibinfo{author}{\bibfnamefont{J.~E.} \bibnamefont{Thomas}},
  \bibinfo{author}{\bibfnamefont{V.}~\bibnamefont{Venturi}},
  \bibinfo{author}{\bibfnamefont{E.}~\bibnamefont{Tiesinga}}, \bibnamefont{and}
  \bibinfo{author}{\bibfnamefont{C.~J.} \bibnamefont{Williams}},
  \bibinfo{journal}{Phys. Rev. A} \textbf{\bibinfo{volume}{66}},
  \bibinfo{pages}{041401} (\bibinfo{year}{2002}).

\bibitem[{\citenamefont{Bruun et~al.}(2005)\citenamefont{Bruun, Jackson, and
  Kolomeitsev}}]{Bruun2005}
\bibinfo{author}{\bibfnamefont{G.~M.} \bibnamefont{Bruun}},
  \bibinfo{author}{\bibfnamefont{A.~D.} \bibnamefont{Jackson}},
  \bibnamefont{and} \bibinfo{author}{\bibfnamefont{E.~E.}
  \bibnamefont{Kolomeitsev}}, \bibinfo{journal}{Phys. Rev. A}
  \textbf{\bibinfo{volume}{71}}, \bibinfo{pages}{052713}
  (\bibinfo{year}{2005}).

\bibitem[{\citenamefont{Pricoupenko}(2006{\natexlab{a}})}]{Pricoupenko2006}
\bibinfo{author}{\bibfnamefont{L.}~\bibnamefont{Pricoupenko}},
  \bibinfo{journal}{Phys. Rev. A} \textbf{\bibinfo{volume}{73}},
  \bibinfo{pages}{012701} (\bibinfo{year}{2006}{\natexlab{a}}).

\bibitem[{\citenamefont{Sakurai and Napolitano}(2010)}]{Sakurai2010}
\bibinfo{author}{\bibfnamefont{J.~J.} \bibnamefont{Sakurai}} \bibnamefont{and}
  \bibinfo{author}{\bibfnamefont{J.~J.} \bibnamefont{Napolitano}},
  \emph{\bibinfo{title}{{Modern Quantum Mechanics}}}
  (\bibinfo{publisher}{Addison-Wesley}, \bibinfo{year}{2010}),
  \bibinfo{edition}{2nd} ed.

\bibitem[{\citenamefont{Braaten and Hammer}(2006)}]{Braaten2006}
\bibinfo{author}{\bibfnamefont{E.}~\bibnamefont{Braaten}} \bibnamefont{and}
  \bibinfo{author}{\bibfnamefont{H.-W.} \bibnamefont{Hammer}},
  \bibinfo{journal}{Phys. Rep.} \textbf{\bibinfo{volume}{428}}
  (\bibinfo{year}{2006}).

\bibitem[{\citenamefont{Efimov}(1970{\natexlab{a}})}]{Efimov1970a}
\bibinfo{author}{\bibfnamefont{V.}~\bibnamefont{Efimov}},
  \bibinfo{journal}{Phys. Lett. B} \textbf{\bibinfo{volume}{33}},
  \bibinfo{pages}{563} (\bibinfo{year}{1970}{\natexlab{a}}).

\bibitem[{\citenamefont{Efimov}(1970{\natexlab{b}})}]{Efimov1970}
\bibinfo{author}{\bibfnamefont{V.~N.} \bibnamefont{Efimov}},
  \bibinfo{journal}{Yad. Fiz.} \textbf{\bibinfo{volume}{12}},
  \bibinfo{pages}{1080} (\bibinfo{year}{1970}{\natexlab{b}}).

\bibitem[{\citenamefont{Amado and Noble}(1972)}]{Amado1972}
\bibinfo{author}{\bibfnamefont{R.~D.} \bibnamefont{Amado}} \bibnamefont{and}
  \bibinfo{author}{\bibfnamefont{J.~V.} \bibnamefont{Noble}},
  \bibinfo{journal}{Phys. Rev. D} \textbf{\bibinfo{volume}{5}},
  \bibinfo{pages}{1992} (\bibinfo{year}{1972}).

\bibitem[{\citenamefont{Efimov}(1973)}]{Efimov1973}
\bibinfo{author}{\bibfnamefont{V.}~\bibnamefont{Efimov}},
  \bibinfo{journal}{Nucl. Phys. A} \textbf{\bibinfo{volume}{210}},
  \bibinfo{pages}{157} (\bibinfo{year}{1973}).

\bibitem[{\citenamefont{Tan}(2012)}]{Tan2012}
\bibinfo{author}{\bibfnamefont{S.}~\bibnamefont{Tan}}, \bibinfo{journal}{Phys.
  Rev. Lett.} \textbf{\bibinfo{volume}{109}}, \bibinfo{pages}{020401}
  (\bibinfo{year}{2012}).

\bibitem[{\citenamefont{Macek and Sternberg}(2006)}]{Macek2006}
\bibinfo{author}{\bibfnamefont{J.~H.} \bibnamefont{Macek}} \bibnamefont{and}
  \bibinfo{author}{\bibfnamefont{J.}~\bibnamefont{Sternberg}},
  \bibinfo{journal}{Phys. Rev. Lett.} \textbf{\bibinfo{volume}{97}},
  \bibinfo{pages}{023201} (\bibinfo{year}{2006}).

\bibitem[{\citenamefont{Macek}(2007)}]{Macek2007}
\bibinfo{author}{\bibfnamefont{J.~H.} \bibnamefont{Macek}},
  \bibinfo{journal}{Nucl. Phys. A} \textbf{\bibinfo{volume}{790}}
  (\bibinfo{year}{2007}).

\bibitem[{\citenamefont{Braaten et~al.}(2012)\citenamefont{Braaten, Hagen,
  Hammer, and Platter}}]{Braaten2012}
\bibinfo{author}{\bibfnamefont{E.}~\bibnamefont{Braaten}},
  \bibinfo{author}{\bibfnamefont{P.}~\bibnamefont{Hagen}},
  \bibinfo{author}{\bibfnamefont{H.-W.} \bibnamefont{Hammer}},
  \bibnamefont{and} \bibinfo{author}{\bibfnamefont{L.}~\bibnamefont{Platter}},
  \bibinfo{journal}{Phys. Rev. A} \textbf{\bibinfo{volume}{86}},
  \bibinfo{pages}{012711} (\bibinfo{year}{2012}).

\bibitem[{\citenamefont{Nishida}(2012)}]{Nishida2012}
\bibinfo{author}{\bibfnamefont{Y.}~\bibnamefont{Nishida}},
  \bibinfo{journal}{Phys. Rev. A} \textbf{\bibinfo{volume}{86}},
  \bibinfo{pages}{012710} (\bibinfo{year}{2012}).

\bibitem[{\citenamefont{Pricoupenko}(2006{\natexlab{b}})}]{Pricoupenko2006a}
\bibinfo{author}{\bibfnamefont{L.}~\bibnamefont{Pricoupenko}},
  \bibinfo{journal}{Phys. Rev. Lett.} \textbf{\bibinfo{volume}{96}},
  \bibinfo{pages}{050401} (\bibinfo{year}{2006}{\natexlab{b}}).

\bibitem[{\citenamefont{Jona-Lasinio et~al.}(2008)\citenamefont{Jona-Lasinio,
  Pricoupenko, and Castin}}]{Jona-Lasinio2008}
\bibinfo{author}{\bibfnamefont{M.}~\bibnamefont{Jona-Lasinio}},
  \bibinfo{author}{\bibfnamefont{L.}~\bibnamefont{Pricoupenko}},
  \bibnamefont{and} \bibinfo{author}{\bibfnamefont{Y.}~\bibnamefont{Castin}},
  \bibinfo{journal}{Phys. Rev. A} \textbf{\bibinfo{volume}{77}},
  \bibinfo{pages}{043611} (\bibinfo{year}{2008}).

\bibitem[{\citenamefont{Hammer and Lee}(2009)}]{Hammer2009}
\bibinfo{author}{\bibfnamefont{H.-W.} \bibnamefont{Hammer}} \bibnamefont{and}
  \bibinfo{author}{\bibfnamefont{D.}~\bibnamefont{Lee}},
  \bibinfo{journal}{Phys. Lett. B} \textbf{\bibinfo{volume}{681}}
  (\bibinfo{year}{2009}).

\bibitem[{\citenamefont{Hammer and Lee}(2010)}]{Hammer2010}
\bibinfo{author}{\bibfnamefont{H.-W.} \bibnamefont{Hammer}} \bibnamefont{and}
  \bibinfo{author}{\bibfnamefont{D.}~\bibnamefont{Lee}}, \bibinfo{journal}{Ann.
  Phys.} \textbf{\bibinfo{volume}{325}}, \bibinfo{pages}{2212}
  (\bibinfo{year}{2010}).

\bibitem[{\citenamefont{Fonseca et~al.}(1979)\citenamefont{Fonseca, Redish, and
  Shanley}}]{Fonseca1979}
\bibinfo{author}{\bibfnamefont{A.~C.} \bibnamefont{Fonseca}},
  \bibinfo{author}{\bibfnamefont{E.~F.} \bibnamefont{Redish}},
  \bibnamefont{and} \bibinfo{author}{\bibfnamefont{P.~E.}
  \bibnamefont{Shanley}}, \bibinfo{journal}{Nucl. Phys. A}
  \textbf{\bibinfo{volume}{320}} (\bibinfo{year}{1979}).

\bibitem[{\citenamefont{Marcelis et~al.}(2008)\citenamefont{Marcelis,
  Kokkelmans, Shlyapnikov, and Petrov}}]{Marcelis2008}
\bibinfo{author}{\bibfnamefont{B.}~\bibnamefont{Marcelis}},
  \bibinfo{author}{\bibfnamefont{S.~J. J. M.~F.} \bibnamefont{Kokkelmans}},
  \bibinfo{author}{\bibfnamefont{G.~V.} \bibnamefont{Shlyapnikov}},
  \bibnamefont{and} \bibinfo{author}{\bibfnamefont{D.~S.}
  \bibnamefont{Petrov}}, \bibinfo{journal}{Phys. Rev. A}
  \textbf{\bibinfo{volume}{77}}, \bibinfo{pages}{032707}
  (\bibinfo{year}{2008}).

\bibitem[{\citenamefont{Nishida et~al.}(2008)\citenamefont{Nishida, Son, and
  Tan}}]{Nishida2008}
\bibinfo{author}{\bibfnamefont{Y.}~\bibnamefont{Nishida}},
  \bibinfo{author}{\bibfnamefont{D.~T.} \bibnamefont{Son}}, \bibnamefont{and}
  \bibinfo{author}{\bibfnamefont{S.}~\bibnamefont{Tan}},
  \bibinfo{journal}{Phys. Rev. Lett.} \textbf{\bibinfo{volume}{100}},
  \bibinfo{pages}{090405} (\bibinfo{year}{2008}).

\bibitem[{\citenamefont{Massignan and Castin}(2006)}]{Massignan2006}
\bibinfo{author}{\bibfnamefont{P.}~\bibnamefont{Massignan}} \bibnamefont{and}
  \bibinfo{author}{\bibfnamefont{Y.}~\bibnamefont{Castin}},
  \bibinfo{journal}{Phys. Rev. A} \textbf{\bibinfo{volume}{74}},
  \bibinfo{pages}{013616} (\bibinfo{year}{2006}).

\bibitem[{\citenamefont{Nishida and Tan}(2010)}]{Nishida2010}
\bibinfo{author}{\bibfnamefont{Y.}~\bibnamefont{Nishida}} \bibnamefont{and}
  \bibinfo{author}{\bibfnamefont{S.}~\bibnamefont{Tan}},
  \bibinfo{journal}{Phys. Rev. A} \textbf{\bibinfo{volume}{82}},
  \bibinfo{pages}{062713} (\bibinfo{year}{2010}).

\bibitem[{\citenamefont{Bethe}(1949)}]{Bethe1949}
\bibinfo{author}{\bibfnamefont{H.}~\bibnamefont{Bethe}},
  \bibinfo{journal}{Phys. Rev.} \textbf{\bibinfo{volume}{76}}
  (\bibinfo{year}{1949}).

\bibitem[{\citenamefont{Madsen}(2002)}]{Madsen2002}
\bibinfo{author}{\bibfnamefont{L.~B.} \bibnamefont{Madsen}},
  \bibinfo{journal}{Am. J. Phys.} \textbf{\bibinfo{volume}{70}}
  (\bibinfo{year}{2002}).

\bibitem[{\citenamefont{Yin et~al.}(2011)\citenamefont{Yin, Zhang, and
  Zhang}}]{Yin2011}
\bibinfo{author}{\bibfnamefont{T.}~\bibnamefont{Yin}},
  \bibinfo{author}{\bibfnamefont{P.}~\bibnamefont{Zhang}}, \bibnamefont{and}
  \bibinfo{author}{\bibfnamefont{W.}~\bibnamefont{Zhang}},
  \bibinfo{journal}{Phys. Rev. A} \textbf{\bibinfo{volume}{84}},
  \bibinfo{pages}{052727} (\bibinfo{year}{2011}).

\bibitem[{\citenamefont{Levy and Keller}(1963)}]{Levy1963}
\bibinfo{author}{\bibfnamefont{B.~R.} \bibnamefont{Levy}} \bibnamefont{and}
  \bibinfo{author}{\bibfnamefont{J.~B.} \bibnamefont{Keller}},
  \bibinfo{journal}{J. Math. Phys.} \textbf{\bibinfo{volume}{4}}
  (\bibinfo{year}{1963}).

\bibitem[{\citenamefont{Gao}(1998)}]{Gao1998a}
\bibinfo{author}{\bibfnamefont{B.}~\bibnamefont{Gao}}, \bibinfo{journal}{Phys.
  Rev. A} \textbf{\bibinfo{volume}{58}}, \bibinfo{pages}{4222}
  (\bibinfo{year}{1998}).

\bibitem[{\citenamefont{Gao}(2009)}]{Gao2009}
\bibinfo{author}{\bibfnamefont{B.}~\bibnamefont{Gao}}, \bibinfo{journal}{Phys.
  Rev. A} \textbf{\bibinfo{volume}{80}}, \bibinfo{pages}{012702}
  (\bibinfo{year}{2009}).

\bibitem[{\citenamefont{Zhang et~al.}(2010)\citenamefont{Zhang, Naidon, and
  Ueda}}]{Zhang2010a}
\bibinfo{author}{\bibfnamefont{P.}~\bibnamefont{Zhang}},
  \bibinfo{author}{\bibfnamefont{P.}~\bibnamefont{Naidon}}, \bibnamefont{and}
  \bibinfo{author}{\bibfnamefont{M.}~\bibnamefont{Ueda}},
  \bibinfo{journal}{Phys. Rev. A} \textbf{\bibinfo{volume}{82}},
  \bibinfo{pages}{062712} (\bibinfo{year}{2010}).

\bibitem[{\citenamefont{Gao}(2011)}]{Gao2011}
\bibinfo{author}{\bibfnamefont{B.}~\bibnamefont{Gao}}, \bibinfo{journal}{Phys.
  Rev. A} \textbf{\bibinfo{volume}{84}}, \bibinfo{pages}{022706}
  (\bibinfo{year}{2011}).

\bibitem[{con()}]{connectiontoGao}
\bibinfo{note}{We can find its connection to Eq.(23) of Ref.~\cite{Gao2009} by
  setting $j=\frac{\sqrt{2}K_l^{c0}}{K_l^{c0}+1}$, and
  $c=\frac{\sqrt{2}}{21}\frac{1+(K_l^{c0})^2}{(1+K_l^{c0})^2}$.}

\bibitem[{Efr()}]{Efremov2013}
\bibinfo{note}{M. A. Efremov, L. Plimak, M, Y. Ivanov, and W. P. Schleich,
  arXiv:1303.5939 (2013).}

\end{thebibliography}

\end{document}